\documentclass{article}
\usepackage{lipsum}
\usepackage{amsmath,amsfonts}
\usepackage{algorithmic}
\usepackage{algorithm}
\usepackage{array}
\usepackage[caption=false,font=normalsize,labelfont=sf,textfont=sf]{subfig}
\usepackage{textcomp}
\usepackage{stfloats}
\usepackage{booktabs}
\usepackage{verbatim}
\usepackage{graphicx}
\usepackage[most]{tcolorbox}
\usepackage{paralist}
\usepackage{xspace}
\usepackage{multirow}
\usepackage{qcircuit}
\usepackage[inkscapelatex=false]{svg}
\usepackage{subcaption}
\usepackage[export]{adjustbox}
\usepackage{url}

\definecolor{codeblue}{rgb}{0.0, 0.0, 0.5}

\makeatletter
\newcommand\codefont{\@setfontsize\customsize{6.5}{7.6}} 
\makeatother

\newcommand{\method}{\ensuremath{\mathtt{QOPS}}\xspace}

\newcommand{\pauliString}{\ensuremath{\mathit{Ps}}\xspace}
\newcommand{\pauliFamily}{\ensuremath{\mathit{F}}\xspace}
\newcommand{\numPauliStrings}{\ensuremath{\mathit{Ps}}\xspace}
\newcommand{\eigenvalues}{\ensuremath{\mathit{EV}}\xspace}
\newcommand{\specifiedOutcome}{\ensuremath{\mathit{M}}\xspace}
\newcommand{\ETO}{\ensuremath{\mathtt{ETO}}\xspace}
\newcommand{\expExpValue}{\ensuremath{\mathit{Exp}}\xspace}
\newcommand{\testCase}{\ensuremath{\mathit{T}}\xspace}
\newcommand{\equivBenchs}{\ensuremath{\mathit{EqBenchs}}\xspace}
\newcommand{\faultyBenchs}{\ensuremath{\mathit{FaultyBenchs}}\xspace}
\newcommand{\zne}[1]{%
\ensuremath{\mathbf{ZNE_{#1}}}%
}

\newcommand{\inlineimage}[1]{%
\raisebox{-.2\height}{\includegraphics[width=1em]{#1}}%
}
\newcommand{\inlineimageb}[1]{%
\raisebox{-.3\height}{\includegraphics[width=0.7em]{#1}}%
}

\newcommand{\ourPS}{\ensuremath{\mathit{PS_{compact}}}\xspace}
\newcommand{\cut}{CUT\xspace}

\lstdefinestyle{smalllisting}{
language=Python, 
basicstyle=\ttfamily\codefont\color{codeblue}, 
keywordstyle=\bfseries\color{blue}, 
columns=fullflexible,commentstyle=\itshape\color{gray}, 
stringstyle=\color{red}, 
showstringspaces=false, 
frame=False, 
breaklines=true, 
}

\usepackage{PRIMEarxiv}

\title{Quantum Program Testing Through Commuting Pauli Strings on IBM's Quantum Computers}

\author{
  Asmar~Muqeet\\
  Simula Research Laboratory \\
  University of Oslo \\
  Oslo\\
  \texttt{asmar@simula.no} \\
   \And
  Shaukat~Ali \\
  Simula Research Laboratory and \\
  Oslo Metropolitan University \\
  Oslo\\
  \texttt{shaukat@simula.no} \\
   \And
  Paolo~Arcaini \\
  National Institute of Informatics \\
  Tokyo\\
  \texttt{arcaini@nii.ac.jp} \\
}

\begin{document}
\maketitle

\begin{abstract}
The most promising applications of quantum computing are centered around solving search and optimization tasks, particularly in fields such as physics simulations, quantum chemistry, and finance. However, the current quantum software testing methods face practical limitations when applied in industrial contexts: (i) they do not apply to quantum programs most relevant to the industry, (ii) they require a full program specification, which is usually not available for these programs, and (iii) they are incompatible with error mitigation methods currently adopted by main industry actors like IBM. To address these challenges, we present \method, a novel quantum software testing approach. \method introduces a new definition of test cases based on Pauli strings to improve compatibility with different quantum programs. \method also introduces a new test oracle that can be directly integrated with industrial APIs such as IBM's Estimator API and can utilize error mitigation methods for testing on real noisy quantum computers. We also leverage the commuting property of Pauli strings to relax the requirement of having complete program specifications, making \method practical for testing complex quantum programs in industrial settings. We empirically evaluate \method on 194,982 real quantum programs, demonstrating effective performance in test assessment compared to the state-of-the-art with a perfect F1-score, precision, and recall. Furthermore, we validate the industrial applicability of \method by assessing its performance on IBM's three real quantum computers, incorporating both industrial and open-source error mitigation methods.
\end{abstract}


\section{Introduction}

Quantum software testing aims to efficiently detect bugs in quantum software to ensure its correctness and reliability~\cite{qse_challenges}. Existing quantum software testing methods adopt classical software testing methods such as search-based testing~\cite{search,mutation2}, mutation testing~\cite{mutation}, combinatorial testing~\cite{Combinatorial}, coverage criteria~\cite{coverage,LongTOSEM2024,quraTestASE23}, property-based testing~\cite{property2}, mutation testing~\cite{mutation,Rui_mutation}, metamorphic testing~\cite{Rui_metamorphic}, and fuzzing~\cite{fuzz}. However, practical limitations exist in applying these existing quantum software testing methods to real quantum computers. 

First, the existing testing methods are incompatible with some types of quantum programs, specifically those performing search and optimization tasks. These quantum programs do not explicitly take inputs, but instead leverage the superposition of all states to find solutions to problems. In contrast, the types of quantum programs supported by existing testing methods require explicit inputs, and their testing involves finding failing inputs. 

Second, most existing quantum testing methods rely on explicit oracles to determine whether a test case passes or fails. These methods often assume that the expected output for each test case is known in advance, which is unrealistic, especially for large quantum programs. As quantum algorithms evolve to handle more complex computations, specifying expected outputs for all potential test cases becomes increasingly impractical.

Third, existing testing methods assume ideal, noise-free quantum computers. However, this is not true for current quantum computers, which are inherently noisy, with errors affecting their computations~\cite{noise}. To mitigate noise errors, the industry has adopted error mitigation methods that can be integrated with optimization programs to solve specific tasks on real quantum computers~\cite{mitiq}. However, current testing methods are not designed to accommodate these error mitigation methods effectively. They often require additional instrumentation to align with error mitigation methods, making them difficult to use with high-level APIs such as IBM's Estimator API for developing quantum programs.

This paper addresses these practical challenges by introducing a novel quantum software testing approach called \method, which utilizes a new test case definition based on \emph{Pauli strings} that does not require explicit inputs for quantum programs, thereby ensuring compatibility with all programs, including optimization and search programs. We also propose a new test oracle that can be integrated with error mitigation methods commonly adopted in the industry. Furthermore, we employ the concept of {\it commuting Pauli strings} and {\it Pauli families} to minimize the number of expected program outputs that must be specified to assess the passing and falling of a test case, making our approach more practical in industrial settings. We evaluated our approach on 194,982 real quantum programs and compared it with the current state-of-the-art test case definition and test oracle~\cite{quitoASE21tool}. The results demonstrate that our approach performs better in test assessment, achieving an F1-score, precision, and recall of 1. Furthermore, we show the industrial relevance of our approach by performing test assessment on IBM's three quantum computers, integrating one industrial error mitigation method and one open-source method. Additionally, we discuss the implications and practicality of our approach and provide insights for practitioners. 

Our key contributions are:
\begin{inparaenum}[1)]
\item A novel testing approach that employs a new test case definition that is compatible with all quantum programs including search and optimization programs.
\item A novel test oracle compatible with error mitigation methods most widely used in industrial settings.
\item An empirical evaluation of test assessment across 194,982 quantum programs, demonstrating the effectiveness of our approach compared to an existing state-of-the-art.
\item An empirical evaluation of test assessment on IBM's three real quantum computers, incorporating both industrial and open-source error mitigation methods to illustrate the practical application of our approach.
\end{inparaenum}

\section{Background}
\subsection{QC Basics and Example} \label{subsec:QC}
\textbf{Qubits:} In classic computing, the basic unit of information is the bit, which can only be in the state of 0 or 1. Quantum computing~(QC), however, uses quantum bits or \textit{qubits}, which can exist in a \textit{superposition} of both $|0\rangle$ and $|1\rangle$ states, with associated amplitudes $(\alpha)$. $\alpha$ is a complex number with both \textit{magnitude} and \textit{phase} in its polar form. A qubit is represented in Dirac notation~\cite{dirac} as: $|\psi\rangle = \alpha_0 |0\rangle + \alpha_1 |1\rangle$, where $\alpha_0$ and $\alpha_1$ are the amplitudes corresponding to states $|0\rangle$ and $|1\rangle$, respectively. The probabilities of a qubit being in state $|0\rangle$ or $|1\rangle$ are determined by the square of the magnitudes of $\alpha_0$ and $\alpha_1$, with the sum of these squared magnitudes equaling 1: $|\alpha_0|^2 + |\alpha_1|^2 = 1$.

\begin{figure}[!tb]
\centering
\includegraphics[width=0.8\columnwidth]{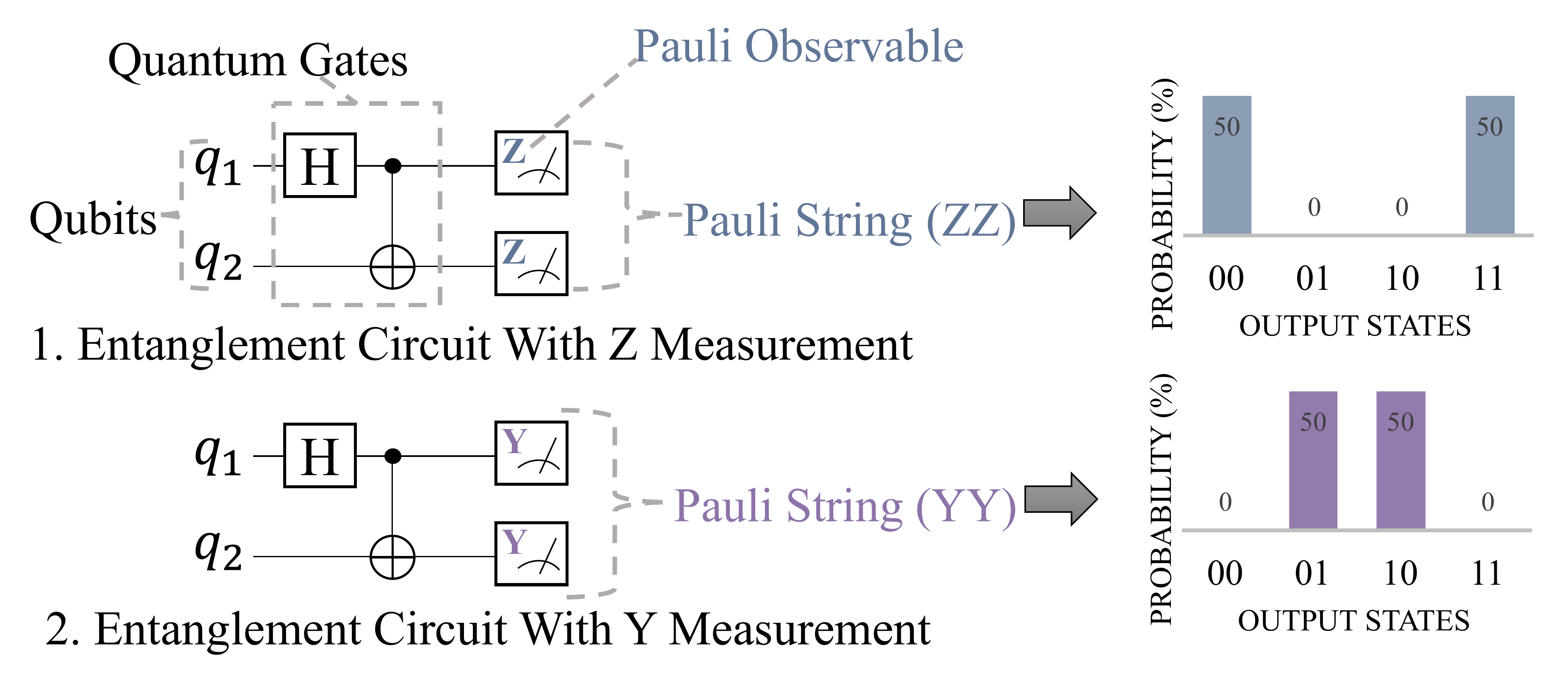}
\caption{Two-qubit entanglement quantum circuit with Z and Y measurement.}
\label{fig:twoqubit}
\end{figure}

\noindent\textbf{Quantum gates:} These unitary operators change a qubit's state based on a unitary matrix~\cite{basic}. For instance, the \textit{NOT} gate flips the state of a qubit. Currently, gate-based quantum computers are programmed as quantum circuits, where the logic is implemented as a sequence of quantum gates applied to qubits. For better readability, we will refer to a quantum program as a ``circuit'' throughout this paper. Figure~\ref{fig:twoqubit} shows two entanglement circuits, each involving two qubits.
Entanglement is a phenomenon where two or more qubits are linked in such a way that the state of one qubit directly affects the state of the other. The circuit starts with two qubits ($q_1$ and $q_2$), initialized to the state $|0\rangle$ by default. First, a \textit{Hadamard} gate ( \inlineimage{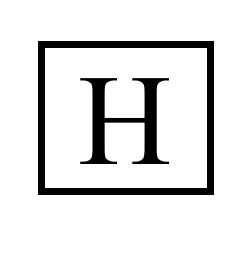} ) is applied to $q_1$, placing it in a superposition of the $|0\rangle$ and $|1\rangle$ states. Next, a \textit{controlled-not} gate ( \inlineimageb{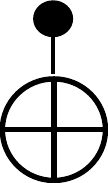} ) is applied. A \textit{controlled-not} gate is a two-qubit quantum gate, where the first qubit ($q_1$) is referred to as the control qubit and the second qubit ($q_2$) as the target qubit. The \textit{controlled-not} gate flips the state of the target qubit if the control qubit is in state $|1\rangle$. The \textit{controlled-not} gate is applied after the \textit{Hadamard} gate to create entanglement between the two qubits.

\noindent\textbf{Measurement and Observable:} Measurement is an operation used to extract the output of a quantum circuit. There are different types of measurement operations, with the type depending on the \emph{observable} being used. Observables are physical quantities that can be measured, such as the position, momentum, and spin of a particle~\cite{Basics}. In QC, observables are represented by Hermitian matrices~\cite{Basics} with predefined eigenvalues, which are used in a measurement operation to extract the outcome of a qubit. Current quantum computers utilize four observables ($Z$, $Y$, $X$, $I$), known as \textit{Pauli observables}, to measure the state of a single qubit~\cite{Basics}. 
For multi-qubit circuits, the measurement operation is represented by the tensor product of single-qubit Pauli observables, referred to as \textit{Pauli strings}. For instance, in Figure~\ref{fig:twoqubit}, the \textit{Z} measurement ( \inlineimage{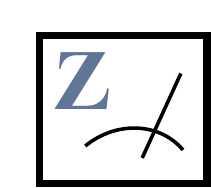} ) in circuit 1 and the \textit{Y} measurement ( \inlineimage{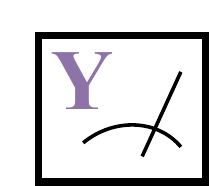} ) in circuit 2 illustrate single-qubit Pauli observables. The overall measurement operation in circuit 1 is defined by Pauli string \textit{ZZ} which is the tensor product of Pauli observable \textit{Z} on qubit $q_1$ and $q_2$. The output of a quantum circuit depends on the Pauli string used for measurement. For example, in Figure~\ref{fig:twoqubit}, the output for Pauli string \textit{ZZ} in circuit 1 is (00) and (11), each with a probability of 0.5. For Pauli string \textit{YY} in circuit 2, the output is (01) and (10), each with a probability of 0.5. For an n-qubit quantum circuit, there are $4^n$ possible combinations of Pauli strings, where $n$ is the number of qubits and 4 represents the choices of Pauli observables ($Z$, $Y$, $X$, $I$) for each qubit.
Pauli strings have {\it eigenvalues} associated with them, which provides information about the possible outcomes of a Pauli string~\cite{basic}. As mentioned before, \textit{Pauli strings} are tensor products of Pauli observables, and \textit{Pauli observables} are Hermitian metrics with predefined eigenvalues. So the eigenvalues of \textit{Pauli strings} are represented by taking the tensor product of eigenvalues of \textit{Pauli observables} that makes up that \textit{Pauli string}. For example Pauli observable \textit{Z} has eigenvalues (1,-1). So the eigenvalues of Pauli string \textit{ZZ} can be calculated by the tensor product of (1,-1) with (1,-1), resulting in (1, -1, -1, 1) as the eigenvalues of Pauli string \textit{ZZ}. The eigenvalues play an important role in the calculation of expectation values (described next) of a Pauli string.

\noindent\textbf{Expectation Value:} The output of a quantum circuit can vary upon repeated measurements due to its inherent probabilistic nature~\cite{basic}. The expectation value is a concept used to describe the average outcome one would anticipate from such repeated measurements. It can be calculated using the formula $E = \sum_{i=0}^{n} w_iP_i$, where $ n $ is the total number of possible output states of a circuit, $ P_i $ is the probability of the $i$th output state, and $w_i$ is the \textit{i}th eigenvalue of the Pauli string. For example, consider the two-qubit circuit in Figure~\ref{fig:twoqubit} with Pauli string \textit{ZZ}. The eigenvalues for \textit{ZZ} are (1, -1, -1, 1). Using the formula and the measurement results from Figure~\ref{fig:twoqubit}, the expectation value for Pauli string \textit{ZZ} is calculated as $ E = (1 * 0.5) + (-1 * 0) + (-1 * 0) + (1 * 0.5) = 1 $.

\subsection{Commuting Pauli Family}\label{sec:paulifam}
\emph{Pauli family} refers to a group of Pauli strings that share a specific property, such as the commuting property~\cite{fastpart}, which we consider in this work. The commuting property states that the product of two Pauli strings is not dependent on the order of application. For example, two Pauli strings \textit{ZZ} and \textit{ZI} commute if $ZZ \cdot ZI = ZI \cdot ZZ$.
As explained in section~\ref{subsec:QC}, for an $n$-qubit quantum circuit, there are $4^n$ Pauli strings, where $n$ is the number of qubits and 4 arises from having four possibilities for each qubit, corresponding to the four types of Pauli observables: $X$, $Y$, $Z$, and $I$. All $4^n$ Pauli strings can be divided into $K$ commuting Pauli families, where $K$ varies for different quantum circuits. Several algorithms exist for generating Pauli families for a quantum circuit, with the state-of-the-art method presented by Reggio et al.~\cite{fastpart}. 
One key advantage of constructing families of commuting Pauli strings is their ability to enable simultaneous measurements. Commuting Pauli strings have the property that the measurement result of a single Pauli string within a family can be used to calculate the expectation values of all Pauli strings within that family~\cite{fastpart, Paulistrings2}. This property of commuting Pauli strings is extensively utilized in current quantum computers for tasks such as error correction and accelerating quantum computations~\cite{Paulistrings2}.

\section{Industrial Context and Challenges}\label{sec:challenge}
In this section, we present the industrial context by highlighting three main challenges associated with applying existing quantum software testing methods from the perspective of IBM's real quantum computers. These challenges are not exclusive to IBM, but also apply to other gate-based quantum computers. However, the observations presented here are from our experience in testing quantum circuits using IBM's quantum computers. The three main challenges are {\it incompatible test cases}, {\it requirement of a complete program specification }, and {\it incompatibility with error mitigation methods}.

\subsection{Incompatible Test Cases}
The first major issue with current quantum software testing methods is that they are not compatible with some types of quantum circuits. Quantum computing's primary application areas, such as physics simulations and quantum chemistry, revolve around optimization and search problems~\cite{application}. The industry is particularly focused on developing quantum circuits that can function on today's noisy quantum computers to address various optimization and search problems. Examples of such circuits are the Quantum Approximate Optimization Algorithm (QAOA) and Grover's algorithm~\cite{VQE,application}. However, the existing quantum software testing methods do not support these optimization and search circuits. The major challenge existing testing methods face for search and optimization circuits is the definition of appropriate test cases. Currently, test cases for quantum circuits are defined in two main ways. The first is {\it binary test cases}, employed by several testing approaches~\cite{quitoASE21tool,search,mutation2,Combinatorial,qoinArxiv}. In this definition, the test cases are represented as binary inputs in which qubits are initialized to either the $|0\rangle$ or $|1\rangle$ state. The second is {\it quantum test cases}, used by other testing methods~\cite{property2,fuzz,quraTestASE23}. Quantum test cases are represented as quantum inputs in which qubits are initialized to a random superposition state $\alpha|0\rangle+\beta|1\rangle$. However, both these test case types are unsuitable for search and optimization circuits like Grover's and QAOA. Indeed, in these circuits, the qubits are initialized to specific superposition states depending on the problem to be solved. For example, in Grover's circuit, the qubits are initialized in an equal superposition of all states, which overwrites any specific initial state set by test cases. This makes traditional test cases unsuitable for testing such circuits. 

It is therefore important to have a suitable test case definition for search and optimization circuits, especially when real quantum computers are predominantly used for quantum simulations and optimization tasks~\cite{application}. We address this challenge by proposing a new test case definition in our approach consisting of Pauli strings that act on the measurement operation rather than
explicitly setting qubit states as traditional test cases do.

\subsection{Requirement of a Complete Program Specification}
The second major issue is the impractical requirement of specifying the expected outputs for all test cases prior to executing the tests. Given that optimization circuits are intended to solve large-scale problems, it is impractical to define the output for all possible test cases. Current quantum software testing methods require explicitly stating whether a test case passes. This is done by comparing the observed test case outcome with the expected test case outcome defined in a program specification. However, creating a comprehensive program specification is challenging, particularly when quantum algorithms are designed to solve computationally intensive problems. For instance, if we only consider binary test cases, a quantum circuit with $n$ qubits can have $2^n$ potential test cases, necessitating an overwhelming number of test cases to be specified. Although it may be feasible to specify the expected output of a few test cases, specifying the output of all test cases is computationally impractical. We reduce the cost of program specification by utilizing the commuting property of Pauli strings described in Section~\ref{sec:paulifam}. Pauli strings can be grouped into commuting Pauli families, allowing one Pauli string from each family to determine the results for the others within that family~\cite{fastpart}. By leveraging this property, the number of outcomes that need to be specified in the program specification can be significantly reduced (more details in Section~\ref{sec:approach}).

\subsection{Incompatibility with Error Mitigation}
The third issue is the assumption made by most current testing methods that quantum computers are ideal and noise-free, which is not the reality. Indeed, real quantum computers exhibit inherent noise, which significantly affects both computational processes and the reliability of circuit outputs~\cite{noise}. However, most current quantum software testing methods assume ideal conditions, making them ineffective in noisy real-world environments. Only three methods~\cite{qoinArxiv,qlearFSE2024,qraft} enable testing on real quantum computers, but these methods come with the cost of training machine learning models, which requires a large amount of data. An alternative approach for testing on real computers could involve using the error mitigation methods proposed in the literature~\cite{qem}. However, these error mitigation methods primarily focus on reducing noise effects on the expectation values of quantum circuits rather than individual outputs, whereas quantum testing methods operate with individual outputs and lack mechanisms or oracles directly addressing expectation values. This creates an incompatibility between current testing methods and error mitigation methods. Furthermore, companies such as IBM have introduced specialized APIs, such as the Estimator API in Qiskit~\cite{qiskit}, to ease calculating expectation values with built-in error mitigation methods. However, current testing methods lack oracles directly addressing expectation values and are incompatible with such APIs. We address this challenge by proposing a new test oracle (Section~\ref{sec:testassessment}) based on expectation values, making it compatible with IBM's Estimator API and other error mitigation methods.

\section{Approach}\label{sec:approach}
\begin{figure}[!tb]
\centering
\includegraphics[width=0.7\columnwidth]{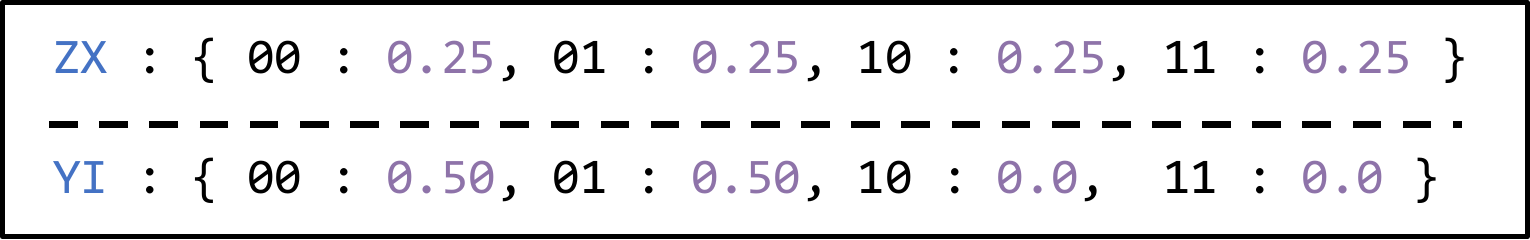}
\caption{An example compact program specification (\ourPS) for the entanglement circuit from Figure~\ref{fig:twoqubit}, showing the outputs of two Pauli strings \textit{ZX} and \textit{YI}. \textit{ZX} belongs to a commuting family consisting of (\textit{XY}, \textit{ZX}, \textit{YZ}) and \textit{YI} belongs to a commuting family consisting of (\textit{IY}, \textit{YY}, \textit{YI}).}
\label{fig:CPS}
\end{figure}

We present an approach named \method: \textbf{Q}uantum pr\textbf{O}gram testing through \textbf{P}auli \textbf{S}trings, addressing the three main challenges outlined in Section~\ref{sec:challenge}. In \method, we utilize Pauli strings for testing quantum circuits.
\method requires two inputs from a tester: the {\it circuit under test} (\cut) and a {\it compact program specification} (\ourPS). In the \ourPS, a tester is required to specify the expected output of any single Pauli string belonging to a commuting Pauli family. The cost of \ourPS is significantly lower than the full program specification required by other testing methods described in Section~\ref{sec:qse}. For example, for a 5-qubit circuit under test, the total number of Pauli strings is $4^5$ (1024). Given that all 1024 Pauli strings can be partitioned into 33 commuting Pauli families, the \ourPS would require a maximum of 33 Pauli strings, which is significantly lower than 1024. An example of \ourPS is shown in Figure~\ref{fig:CPS}, which specifies the outputs of two Pauli strings \textit{ZX} and \textit{YI} for the entanglement circuit in Figure~\ref{fig:twoqubit}.
In the next subsections, we first provide the definition of our test cases and the test oracle along with their rationale in Section~\ref{sec:testcase}. The overall working of the approach is described in Section~\ref{sec:overview}.

\subsection{Test Case and Oracle}~\label{sec:testcase}
\begin{figure}[!tb]
\includegraphics[width=\columnwidth]{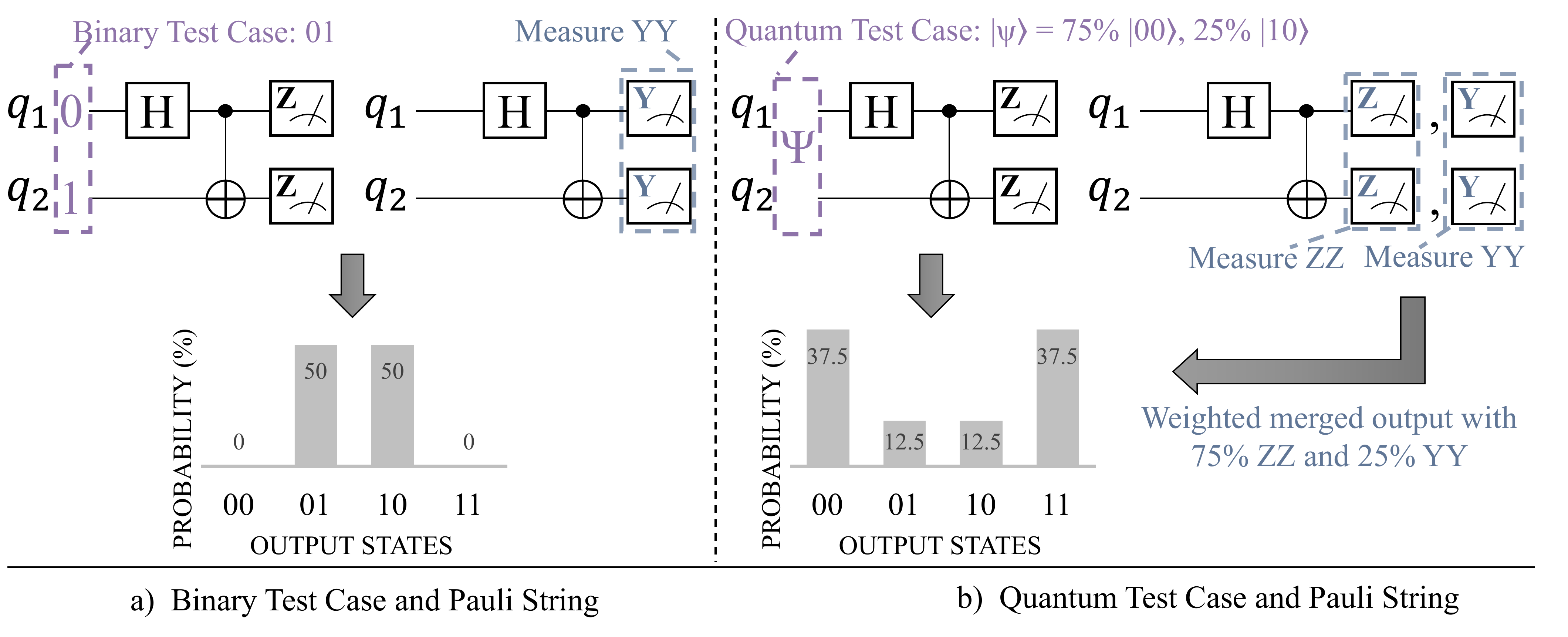}
\caption{Example comparison of traditional test case definition (Binary, Quantum) with Pauli Strings.}
\label{fig:example}
\end{figure}

\method employs a new definition of test case consisting of Pauli strings that act on the measurement operation rather than explicitly setting qubit states as traditional test cases do. We believe that different Pauli strings can identify the same faults in a circuit that are captured by explicitly setting quantum states. The rationale is that traditional test cases set the initial state of a circuit to observe a specific output. However, we noticed that varying the measurement operation with different Pauli strings can also produce the same output. In Figure~\ref{fig:example}, we show an intuitive example by considering the entanglement circuit as the circuit under test.
Figure~\ref{fig:example}a shows the execution result of a binary test case $01$, which sets qubit $q_1$ to $|0\rangle$ and $q_2$ to $|1\rangle$ for the test circuit. However, we can also observe the same execution result of the binary test case by using Pauli string \textit{YY} as shown in Figure~\ref{fig:example}a and not setting the qubit states explicitly. Figure~\ref{fig:example}b shows a quantum test case execution result where the quantum state $\Psi$ is set in a superposition of $|00\rangle$ and $|10\rangle$ with 75\% and 25\% probability for each state, respectively. We can achieve the same output result as of the quantum test case, if we execute the quantum circuit twice, once with the Pauli string \textit{ZZ} and second with the Pauli string \textit{YY} and merge the two outputs by a weighted sum of both distributions, with 75\% weight for \textit{ZZ} and 25\% weight for \textit{YY}. This process of merging the results of different Pauli strings is commonly used in quantum optimization circuits~\cite{VQE}. This indicates that merging outputs of different Pauli strings with different weights could be analogous to quantum test cases without explicitly setting the state of qubits. However, the combination of Pauli strings that are analogous to a specific quantum input for a specific circuit could be different for different circuits. The advantage of using Pauli strings as tests over traditional test cases is that they are associated with the measurement operation, which is independent of the initialization logic of a circuit, making them suitable for optimization and search circuits.

\textbf{Test case definition:} We define our test cases based on the formulation $\testCase = \sum_{i=0}^{S} c_i\pauliString_i$, where $S$ is a subset of Pauli strings chosen randomly from a particular Pauli family, $\pauliString_i$ is the $i$th Pauli string in $S$, and $c_i$ is the weight assigned to the $i$th Pauli string. For example, consider the subset $S$ consisting of \textit{IZ}, \textit{ZI}, and \textit{ZZ}. Then, a test case \testCase can be written as $\testCase = c_1IZ + c_2ZI + c_3ZZ$. Here, $c_1$, $c_2$, and $c_3$ are the weights assigned to the Pauli strings to define the contribution of each Pauli string in the overall output. We use the weighted sum format to define our test case because we hypothesize that it is analogous to the traditional quantum test cases. For example, we show in Figure~\ref{fig:example}b, that the output obtained by executing a quantum test case $\Psi$ can also be obtained by executing the Pauli strings \textit{ZZ} and \textit{YY} and performing a weighted sum of their outputs. Therefore, we can represent the same quantum test case $\Psi$ in our Pauli string test case definition as $\testCase = 0.75ZZ + 0.25YY$.

\textbf{Test Oracle:} With our test case definition, we can randomly generate many test cases. However, we also need an oracle to determine whether the circuit under test (\cut) has failed a test case. For this, we define the oracle \textit{Expected Expectation} (\expExpValue), which is the total expectation value of a test case \testCase after execution. This value is calculated by $\sum_{i=0}^{\numPauliStrings} c_iE_i$, where \numPauliStrings is the total number of Pauli strings in test case \testCase, $E_i$ is the expectation value of the $i$th Pauli string in test case \testCase, and $c_i$ is the weight associated with the $i$th Pauli string in test case \testCase. Using the expectation value formula from Section~\ref{subsec:QC}, we can transform the equation as
\begin{equation}\label{eq:exp1}
\expExpValue = \sum_{i=0}^{\numPauliStrings} c_iE_i = \sum_{i=0}^{\numPauliStrings} c_i\sum_{j=0}^{\pauliString_i} w_jP_j
\end{equation}
where $\pauliString_i$ is the $i$th Pauli string in test case \testCase, $P_j$ is the probability of the $j$th output state in the output of Pauli string $\pauliString_i$, and $w_j$ is the $j$th eigenvalue of the $i$th Pauli string. Our test case consists of commuting Pauli strings, and, therefore, the output of a single Pauli string can be used to calculate the expectation value of other Pauli strings. By utilizing this concept, we can simplify Equation~\ref{eq:exp1} as
\begin{equation}\label{eq:exp2}
\expExpValue = \sum_{(i,j)\in I} c_iw_{i,j}M_j
\end{equation}
where $I = \{(i,j) | 0 \leq i \leq \numPauliStrings$ and $0 \leq j \leq \pauliString_i\}$, \numPauliStrings is the total number of Pauli strings in test case \testCase, $\pauliString_i$ is the $i$th Pauli string in the test case \testCase, $c_i$ is the weight of the $i$th Pauli string in test case \testCase, $w_{i,j}$ is the $j$th eigenvalue of the $i$th Pauli string in test case \testCase, and $M_j$ is the value of the $j$th output state of the Pauli string result given in the \ourPS.

For example, consider the test case $\testCase = 0.3ZI + 0.4IZ$; the output of Pauli string $ZZ$ is specified in \ourPS as (00, 01, 10, 11) with a probability of (0.5, 0.2, 0, 0.3) respectively. The eigenvalues of $ZZ$ are (1,-1,-1, 1). Given that \textit{ZZ}, \textit{ZI}, and \textit{IZ} belong to the same commuting family, we can calculate the total expectation value of test case \testCase by utilizing the output of \textit{ZZ}. Substituting the values in Equation~\ref{eq:exp2}:
\begin{equation*}
\expExpValue = (c_0w_{0,0}M_0 + c_0w_{0,1}M_1 + c_0w_{0,2}M_2 + c_0w_{0,3}M_3) 
\end{equation*}
\begin{equation*}
\:\quad+\:(c_1w_{1,0}M_0 + c_1w_{1,1}M_1 + c_1w_{1,2}M_2 + c_1w_{1,3}M_3) 
\end{equation*}
\begin{equation*}
\expExpValue = (0.3*1*0.5 + 0.3*-1*0.2 + 0.3*-1*0 + 0.3*1*0.3) 
\end{equation*}
\begin{equation*}
\:\quad+\:(0.4*1*0.5 + 0.4*-1*0.2 + 0.4*-1*0 + 0.4*1*0.3)=1.56
\end{equation*}
The expected expectation value of the test case \testCase would be 1.56.

\subsection{Overview}\label{sec:overview}
\begin{figure}[!tb]
\centering
\includegraphics[width=0.8\columnwidth]{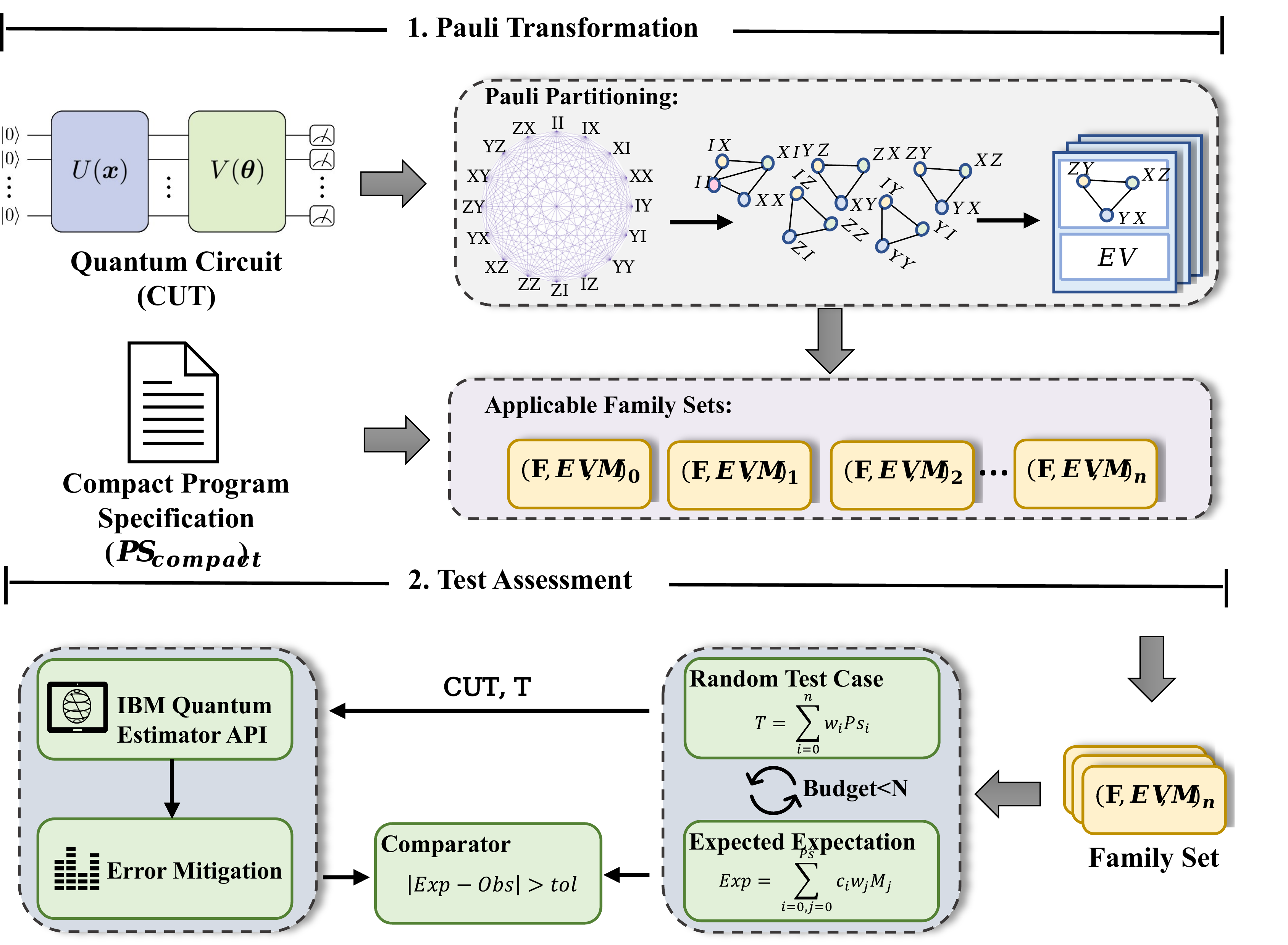}
\caption{Overview of \method: The tuple $(\pauliFamily, \eigenvalues, \specifiedOutcome)$ includes the Pauli family (\pauliFamily), the eigenvalues (\eigenvalues) of the Pauli strings in family \pauliFamily, and the Specified outcome (\specifiedOutcome) of a single Pauli string in \pauliFamily based on a compact program specification (\ourPS).}
\label{fig:overview}
\end{figure}

Figure~\ref{fig:overview} shows the overview of \method approach consisting of two main parts, {\it Pauli Transformation}, and {\it Test Assessment}, which are further described in the following subsections.
\subsubsection{Pauli Transformation}
\method first generates families of commuting Pauli strings that can be used for testing a quantum circuit by taking \cut and \ourPS as inputs. To create Pauli families, \method employs the partitioning algorithm developed by Reggio et al.~\cite{fastpart}, details of which can be found in their respective work. Here, we summarize the partitioning algorithm's operation in two main steps. First, the algorithm starts by identifying a set of Pauli strings that fulfills the commuting condition. Next, for a given \cut and the selected set of Pauli strings, it finds a matrix that diagonalizes all the Pauli strings in the selected set. If such a matrix is successfully found, the commuting set is considered one family, and the process is repeated until no more sets can be found. 
The output of the partitioning algorithm is a list of $K$ Pauli families and the eigenvalues of each Pauli string within a family. Table~\ref{tab:partition} shows the output of the partitioning algorithm for the entanglement circuit from Figure~\ref{fig:twoqubit} taken as \cut.
\begin{table}[!tb]
\caption{Pauli Families with corresponding eigenvalues identified in Pauli Transformation stage for Entanglement circuit in Figure~\ref{fig:twoqubit} taken as \cut.}
\label{tab:partition}
\centering
\resizebox{0.55\columnwidth}{!}{%
\begin{tabular}{c|c}
\toprule
\textbf{Family} & \textbf{EigenValues} \\
\midrule
II,IZ,ZI,ZZ & (1, 1, 1, 1), (1, -1, 1, -1), (1, 1, -1, -1) , (1, -1, -1, 1) \\
IX,XI,XX & (1, -1, 1, -1), (1, 1, -1, -1), (1, -1, -1, 1) \\
IY,YI,YY & (-1, 1, -1, 1), (-1, -1, 1, 1), (1, -1, -1, 1) \\
XY,ZX,YZ & (-1, 1, -1, 1), (-1, -1, 1, 1), (-1, 1, 1, -1) \\
XZ,YX,ZY & (-1, 1, -1, 1), (-1, -1, 1, 1), (-1, 1, 1, -1) \\ 
\bottomrule
\end{tabular}
}
\end{table}
The column Family shows the identified Pauli Families and the column EigenValues shows the eigenvalues associated with each Pauli string within a Pauli family. Given the measurement result of any Pauli string in a family, these eigenvalues can be utilized to calculate the test oracle for a test case consisting of any subset of Pauli string within the family as shown in Section~\ref{sec:testcase}.

Once \method has identified the Pauli families and the corresponding eigenvalues, it checks the \ourPS to filter out non-relevant families for testing. Depending on the number of Pauli strings specified in the \ourPS, not all families may be suitable for inclusion in the testing process. Ideally, the \ourPS should specify at least one Pauli string for each Pauli family to ensure all families can be effectively utilized during testing. However, depending on the number of Pauli strings that can be specified by the tester, the number of applicable Pauli families are reduced accordingly. As shown in Figure~\ref{fig:overview}, the result of the Pauli Transformation stage in \method is an applicable set of families. Each item in this set is represented as a tuple containing three values: \textbf{\pauliFamily} - a specific Pauli family, \textbf{\eigenvalues} - the corresponding eigenvalues of the family, and \textbf{\specifiedOutcome} - the measurement outcome specified in \ourPS for any Pauli string from the family \textbf{\pauliFamily}.

\subsubsection{Test Generation and Assessment}\label{sec:testassessment}
The second part of \method aims to test the quantum circuit by utilizing the applicable family set from the Pauli Transformation stage. As shown in Figure~\ref{fig:overview}, for each family in the applicable family set, \method generates random test cases along with their expected outcome for a specific budget. The user defines the budget, and decides how many test cases to generate for each family. In our experiments, we set the budget to 1000. As shown in Figure~\ref{fig:overview}, the test assessment follows a simple process. For each family, a number of random test cases are generated depending on the budget, and the oracle for each test case is automatically calculated using the \ourPS. The test cases and the \cut are then passed to IBM's Estimator API to execute the \cut with the test case and obtain the results. In case of error mitigation enabled, the result is first parsed to the error mitigation methods integrated in IBM's Estimator API or any open source error mitigation method. The observed results are then compared with the oracle using a threshold value set by the tester. If the difference between the observed result and the oracle exceeds the threshold, the test case is considered failed, and the fault is reported to the tester.

\subsubsection{Integration with Error Mitigation}
\begin{figure}[!tb]
\centering
\includegraphics[width=0.85\columnwidth]{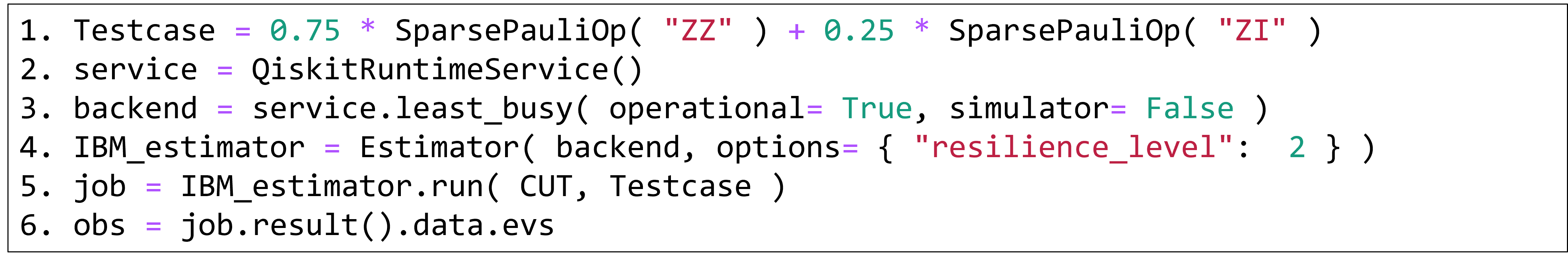}
\caption{An example of IBM Estimator API to execute the test case with inbuilt error mitigation enabled.}
\label{fig:estimator}
\end{figure}

\method's test case definition and oracle are directly compatible with error mitigation methods as we operate with the expectation value and not with individual outputs. Figure~\ref{fig:estimator} illustrates an example of test case executed with IBM's Estimator API, using the zero noise extrapolation (ZNE) technique for error mitigation.
In Line 1, we define a test case as $\testCase = 0.75*ZZ + 0.25*ZI$. In Lines 2-3, we select an IBM quantum computer that is least busy at the moment. Line 4 invokes the IBM Estimator API with the resilience level set to 2, enabling the inbuilt ZNE error mitigation technique. Finally, in Line 5, we submit our test case along with the \cut to the IBM quantum computer. Note that our test case definition is directly compatible, so we can pass the defined test case variable directly to the Estimator API. In Line 6, we obtain the error-mitigated observed expectation value of the test case, which we then compare with our own calculated expected expectation value for assertion.

\section{Evaluation and Analysis}
\subsection{Research Question}
We assess the effectiveness of \method in detecting faults in quantum circuits by answering the following research questions (RQs):
%
\begin{itemize}
\item[\textbf{\hspace{0.4em}RQ1}] How comparable is \method to the most widely used test case definition in terms of capturing faults in quantum circuits?
\item[\textbf{\hspace{0.4em}RQ2}] How effective is \method when applied to testing quantum circuits on real noisy quantum computers?
\end{itemize}

\subsection{Experiment Design}\label{sec:experiment}

\subsubsection{Benchmarks}
We employ the MQT benchmark~\cite{mqt}, a repository of diverse quantum circuits with real-world applications for evaluation of \method. MQT includes 18 different quantum algorithms implemented with varying qubit sizes. We selected all 18 algorithms, but limited the number of qubits to between 2 and 10 to ensure reasonable execution times for the quantum simulator and real computers. This selection resulted in 157 quantum circuits for testing. Additionally, we created faulty versions of these 157 circuits using Muskit~\cite{mutation}, applying all possible mutation operators defined in Muskit. This resulted in 194,982 quantum circuits, comprising 30,084 equivalent circuits and 164,898 faulty circuits, which were used to evaluate \method. We benchmarked \method against the most widely used test case definition, the binary test cases, and the common test oracle based on the chi-square statistical test. These are prevalent in current testing methods~\cite{quitoASE21tool,search,mutation2,Combinatorial,qoinArxiv}. For better readability, we abbreviate the existing test case definition and oracle as \ETO for the rest of the paper.

\subsubsection{Experiment Settings}\label{sec:settings}
The implementation of \method and the experiment results are publicly available in our repository~\cite{sourcecode}. We use the Qiskit framework for quantum circuit execution~\cite{qiskit}. To automatically obtain the full program specifications for the benchmarks for \ETO, we use Qiskit's AER simulator to get the correct circuit outputs. We also use Qiskit's AER simulator to get compact program specifications \ourPS for \method, specifying only one Pauli string for each commuting family for a specific quantum circuit. For execution on real quantum computers, we use IBM's three publicly accessible quantum computers (IBM-Brisbane, IBM-Osaka, and IBM-Kyoto) and employ IBM's Estimator API for circuit execution. For the number of test cases, we set \method budget to 1000 test cases per Pauli family, whereas for \ETO we used all possible test cases, which results in $2^n$ test cases, where $n$ is the number of qubits in the circuit under test. 

Regarding error mitigation on real computers, the most common methods are probabilistic error cancellation (PEC), zero-noise extrapolation (ZNE), and Clifford data regression (CDR)~\cite{mitiq}. IBM's Estimator API provides built-in error mitigation based on ZNE and PEC. Given the exponential cost of PEC~\cite{peccost} and IBM's monthly limit of 10 minutes for using real quantum computers, we selected the ZNE method as the error mitigation method in the Estimator API. Additionally, we used an open-source version of the CDR method~\cite{mitiq} for error mitigation to demonstrate \method's applicability with both industrial and open-source error mitigation methods.

\method requires a tolerance threshold to assert whether a test case fails. For a fair comparison between \method and \ETO, we set this threshold to the same value used in the statistical test assertion in \ETO, which is 1\% for evaluation on the quantum simulator. For real quantum computers, considering that error mitigation methods have their own error margins, we used the Standard Error~\cite{stderr} as the threshold. To calculate the standard error for a quantum circuit, we executed the non-faulty version 10 times with error mitigation and calculated the difference between the ideal value and the error-mitigated value for each run. The standard error was then calculated using the formula $SE = \frac{\sigma}{\sqrt{n}}$, where $\sigma$ is the standard deviation of the differences between the ideal and error-mitigated values over 10 runs, and $n$ is the total number of runs, which is 10. The standard error value varies depending on the margin of error of an error mitigation method. Thus, it is different for each quantum circuit executed with a specific error mitigation method.

\subsubsection{Metrics}
For RQ1, we compare the test assessment results of \method with those of \ETO using a classification problem framework. In this context, a method's accuracy in distinguishing between faulty circuits and equivalent or non-faulty circuits indicates its effectiveness. Therefore, we compare \method with \ETO based on commonly used quality metrics: F1-score, precision, and recall~\cite{commonf3}. These metrics are calculated using their standard formulas~\cite{commonf3}. We compute F1-score, precision, and recall separately for equivalent and faulty versions of quantum circuits, comparing the results with the ground truth for both \method and \ETO. The ground truth is calculated based on the Hellinger distance~\cite{Hellinger2} between the original circuit and the mutant circuit generated by Muskit. Hellinger distance is a commonly used metric for checking the equivalence of two quantum circuits based on their outputs~\cite{Hellinger2}. All circuits that result in a zero value for Hellinger distance are classified as equivalent circuits, and others are classified as faulty circuits. For quality metric calculation, we define True Positive (TP), False Negative (FN), False Positive (FP), and True Negative (TN), in comparison to ground truth as follows: 
\textbf{TP} - Both the ground truth and \ETO's (\method's) test assessment say ''faulty'';
\textbf{FP} - \ETO's (\method's) test assessment says ''faulty'', but the ground truth says ''not faulty''; 
\textbf{FN} - \ETO's (\method's) test assessment says ''not faulty'', but the ground truth says ''faulty''; and 
\textbf{TN} - Both the ground truth and \ETO's (\method's) test assessment say ``not faulty''.

For RQ2, since \ETO is incompatible with error mitigation methods, we can only evaluate \method on real quantum computers. To demonstrate \method's applicability on real computers, we use the same quality metrics as in RQ1. We then analyze how the results differ from our findings in RQ1 when \method is integrated with error mitigation methods and evaluated on real quantum computers.

\subsection{Results}
\subsubsection{RQ1--Comparison with \ETO}
Table~\ref{tab:RQ1} compares \ETO with \method in terms of F1-score, precision, and recall. The columns labeled \equivBenchs and \faultyBenchs present the test assessment results for equivalent and faulty circuits for both \ETO and \method when compared against the ground truth. Table~\ref{tab:RQ1} shows that \method outperforms \ETO in accurately identifying faulty and equivalent circuits without any false positives or false negatives. Although \ETO generally performed well, it exhibited higher false positives and false negatives for some equivalent and faulty circuits compared to \method. Specifically, \ETO misclassified five out of 30,084 equivalent circuits, likely due to the chi-square statistical test's inherent margin of error for false positives. Regarding faulty circuits, \ETO failed to detect faults in 484 out of 164,898 cases. These undetected faults predominantly involved a specific type of fault known as a phase flip fault~\cite{phaseflip}. Detecting phase flip faults is particularly challenging, because it requires altering the Pauli observables in the measurement operation. The chi-square test employed by \ETO is less effective in identifying this type of fault, leading to a higher number of undetected cases. On the other hand, \method proves to be highly effective in capturing phase flip faults. This effectiveness comes from the fact that \method utilizes test cases that consist of Pauli strings, where each Pauli string is composed of different Pauli observables. Overall, Table~\ref{tab:RQ1} shows that \method provides better accuracy in terms of identifying faulty circuits and equivalent circuits with no false positives and false negatives.
\begin{table}[!tb]
\caption{RQ1 -- Quality metric comparison between \ETO and \method against the ground truth. The total number of circuits evaluated in the \equivBenchs category is 30,084, and in the \faultyBenchs category is 164,898.}
\label{tab:RQ1}
\setlength{\tabcolsep}{3pt}
\centering
\resizebox{0.5\columnwidth}{!}{%
\begin{tabular}{c|cc|cc}
\toprule
\multirow{2}{*}{\textbf{Metric}} & \multicolumn{2}{c|}{\textbf{\ETO}} & \multicolumn{2}{c}{\textbf{\method}} \\ \cline{2-5} 
& \multicolumn{1}{c|}{\textbf{\equivBenchs}} & \textbf{\faultyBenchs} & \multicolumn{1}{c|}{\textbf{\equivBenchs}} & \textbf{\faultyBenchs} \\
\midrule
F1-Score & 0.992 & 0.999 & \textbf{1.000} & \textbf{1.000} \\
Precision & 0.984 & 1.000 & \textbf{1.000} & \textbf{1.000} \\
Recall & 1.000 & 0.997 & \textbf{1.000} & \textbf{1.000} \\ 
\bottomrule
\end{tabular}
}
\end{table}
\begin{tcolorbox}[colback=blue!5!white,colframe=white,breakable]
\textbf{RQ1:} Compared to state-of-the-art, \method is more effective in test assessment with a perfect F1-score, precision, and recall, indicating that \method is a more reliable and alternative approach for testing quantum circuits. 
\end{tcolorbox}

\subsubsection{RQ2--Integration with error mitigation}
RQ2 assesses whether \method can be integrated with error mitigation methods and successfully capture faults in quantum circuits on real quantum computers. Given that IBM's quantum computers have a monthly usage limit of 10 minutes, executing all 194,982 circuits on real computers was not possible. To fit within these usage limits, we reduced the number of circuits from RQ1. We employed the Fisher-Jenks natural break algorithm~\cite{fisher} to identify representative faulty circuits for the 157 original circuits, based on the diversity in the \method oracle results. The Fisher-Jenks algorithm is a widely used method for automatically selecting a representative subset from large datasets based on diversity criteria~\cite{fisher2}. The diversity criterion was the expected expectation value of the failed test cases from the RQ1 experiments. This algorithm resulted in 10 faulty circuits per original circuit, yielding 1,570 circuits to be evaluated on real quantum computers. For error mitigation methods, as stated in Section~\ref{sec:settings}, we used both ZNE and CDR methods. IBM provides four different configurations of the ZNE method: Linear (L), Cubic (C), Quartic (Q), and Richardson (R), and we treated each configuration as a separate method. This resulted in five different error mitigation methods, which we evaluated on IBM's three real quantum computers.

Table~\ref{tab:RQ2} shows the results of evaluating 1,570 faulty circuits on the IBM's three real quantum computers: IBM-Brisbane, IBM-Osaka, and IBM-Kyoto.
\begin{table*}[!tb]
\caption{RQ2 -- Quality metric comparison between different error mitigation methods integrated with \method against the ground truth evaluated on IBM's real quantum computers. ZNE configurations are also presented with initials as L-Linear, C-Cubic, Q-Quartic, and R-Richardson. }
\label{tab:RQ2}
\setlength{\tabcolsep}{4pt}
\resizebox{1\textwidth}{!}{%
\begin{tabular}{c|ccccc|ccccc|ccccc}
\toprule
& \multicolumn{5}{c|}{IBM-Brisbane} & \multicolumn{5}{c|}{IBM-Osaka} & \multicolumn{5}{c}{IBM-Kyoto}\\
\cmidrule{2-16}
\textbf{Metric} & \textbf{CDR} & \zne{L} & \zne{C} & \zne{Q} & \zne{R} & \textbf{CDR} & \zne{L} & \zne{C} & \zne{Q} & \zne{R} & \textbf{CDR} & \zne{L} & \zne{C} & \zne{Q} & \zne{R} \\
\midrule
\textbf{Precision} & 1.000 & 1.000 & 1.000 & 1.000 & 1.000 & 1.000 & 1.000 & 1.000 & 1.000 & 1.000 & 1.000 & 1.000 & 1.000 & 1.000 & 1.000 \\
\textbf{Recall} & 0.986 & 0.999 & 0.996 & 0.985 & 0.998 & 0.983 & 0.998 & 0.993 & 0.986 & 0.999 & 0.982 & 0.999 & 0.992 & 0.985 & 0.998 \\
\textbf{F1 score} & 0.993 & \textbf{1.000} & 0.998 & 0.993 & 0.999 & 0.991 & 0.999 & 0.997 & 0.993 & \textbf{1.000} & 0.991 & \textbf{1.000} & 0.996 & 0.992 & 0.999 \\ 
\textbf{False Negatives} & 22.00 & \textbf{1.000} & 5.000 & 15.00 & 2.000 & 24.00 & 2.000 & 8.000 & 14.00 & \textbf{1.000} & 27.00 & \textbf{1.000} & 10.00 & 16.00 & 3.000 \\

\bottomrule
\end{tabular}
}
\end{table*}
The columns display the F1 score, precision, and recall for each error mitigation method and computer pair, calculated by comparing the results with the ground truth. From Table~\ref{tab:RQ2}, we observe that \method performed well across all computer and error mitigation method pairs. The variation in F1 scores for each pair is attributed to the differing levels of noise in each computer and the varying effectiveness of error mitigation methods for different circuits. Notably, the best error mitigation method integrated with \method is the ZNE linear configuration (\zne{L}), which had only one false negative on IBM-Bisbane and IBM-Kyoto and two false negatives on IBM-Osaka. The second best method is the ZNE Richardson (\zne{R}) method. Conversely, the worst method is CDR, with over 20 false negatives across all computers. 

Each error mitigation method has its own time cost for filtering noise from the circuit output. Thus, choosing an appropriate error mitigation method is important. Table~\ref{tab:RQ2} indicates that IBM's integrated ZNE method outperforms the CDR method, particularly with its linear configuration. Overall, \method can successfully perform test assessments on real computers when integrated with error mitigation methods. Testers using IBM's APIs can conduct testing activities on real computers without additional instrumentation for integrating error mitigation with \method. However, the performance of \method is dependent on the accuracy of the error mitigation methods.
\begin{tcolorbox}[colback=blue!5!white,colframe=white,breakable]
\textbf{RQ2:} Compared to state-of-the-art, \method is directly compatible with error mitigation methods, allowing \method to successfully perform test assessments on real computers using APIs with integrated error mitigation methods such as IBM's Estimator API. 
\end{tcolorbox}

\subsection{Threats to validity}
\textbf{Construct Validity}~\cite{Wohlin2012} refers to how well a measurement evaluates the intended theoretical concept. One threat to construct validity involves the metrics used to assess the effectiveness of \method. A testing method must accurately differentiate between faulty and non-faulty circuits. The effectiveness of one testing method compared to another depends on the number of false positives and false negatives produced during the assessments. This work uses F1-score, precision, and recall as performance metrics. Low precision and recall indicate higher false positives and negatives, while the F1-score measures the overall performance of a test assessment method. Another concern is the selection of error mitigation methods and the real computers used. Different quantum computers exhibit varying noise levels, and different error mitigation methods have different error margins for different circuits executed on different computers. In this paper, we focused on all of IBM's quantum computers that were publicly available during the experiment. We concentrated on error mitigation methods commonly used in the industry. IBM provides two error mitigation methods: ZNE and PEC. Given our limited access to IBM's quantum computers, the only feasible error mitigation method was ZNE. Additionally, we used an open-source error mitigation method, CDR, to demonstrate the applicability of \method with both industrial and open-source error mitigation methods.

\textbf{Internal Validity}~\cite{Wohlin2012} concerns how experiments can establish a causal relationship between independent and dependent variables. One threat to internal validity involves the tolerance parameter of \method. The tolerance parameter determines the acceptable error margin when comparing the observed expectation value of the test cases with the expected expectation. This parameter depends on the circuit and requires domain knowledge to be set correctly. In our experiments, to ensure a fair comparison with the \ETO and \method, we chose the tolerance value to be the same as the threshold used in \ETO's statistical test for passing and failing a hypothesis, which is 1\%. Although using a different tolerance value could affect the outcome, we found in the RQ1 results that a tolerance value of 1\% was optimal, as it resulted in zero false positives and negatives for \method. For evaluations on real computers, since error mitigation methods have their own error margins that vary for different circuits, a single tolerance value cannot be established. To address this, we used 1\% plus the standard error value of the original circuit for 10 consecutive runs executed with error mitigation as the threshold value. The standard error is calculated automatically for each circuit and error mitigation pair before performing the full test assessment.

\textbf{Conclusion validity}~\cite{Wohlin2012} focuses on the statistical significance of the results derived from an experiment.
One threat to conclusion validity is using random test cases in the test assessment by \method. Using random test cases does not guarantee that \method will perform consistently if executed multiple times. However, the main objective of this study is to assess whether \method can identify all the faults. Therefore, we focused on evaluating a large number of circuits, which limited our ability to run the experiment multiple times. Additionally, we set the number of test cases for each Pauli family to 1000 to increase the probability of finding a test case that captures the fault in a faulty circuit. The RQ1 results indicate that \method, even with random test cases, could correctly identify all the faulty circuits. Multiple runs to reduce random bias are only necessary when there is a chance that a better solution exists. However, in our case, a single experiment run was sufficient to find the best solution, reducing the need for multiple experiment evaluations.

\textbf{External Validity}~\cite{Wohlin2012} refers to how our method can be applied to other datasets and domains. A challenge in this area is the selection of quantum circuits for the test assessment, as a diverse set of circuits would result in more generalizable outcomes. To mitigate this issue, we used the widely accepted benchmark MQT~\cite{mqt} to choose circuits representing the most commonly utilized quantum circuits. Another potential threat is the number of qubits used, as increasing the number of qubits directly adds to the complexity of quantum circuits. Current quantum simulators can only handle up to 10 qubits in a reasonable time~\cite{speed}. In this study, our goal was to test many circuits for generalizability, which restricted the number of qubits we could use, as more qubits significantly increase the execution time of circuits on simulators. Thus, to evaluate a large number of circuits within a practical time frame, we set the maximum number of qubits to 10.

\section{Discussion}
\subsection{Practical Implications}
This study presented a new quantum circuit testing approach (\method) to solve three practical challenges to ensure the reliable development of QC solutions in an industrial context, particularly for IBM's real quantum computers. Indeed, the incompatibility of current quantum software testing methods with circuits designed for optimization and search tasks limits their practical utility. Since the primary industrial application of QC is in optimization and search problems, this gap presents a substantial barrier. Additionally, the impractical requirement to specify expected outputs for all test cases before execution, challenges the scalability and practicality of current testing methods in real-world scenarios. Furthermore, the assumption that quantum computers are ideal and noise-free fails to reflect their actual state. In reality, quantum computers are susceptible to noise, which affects their performance and accuracy.

To address these practical challenges, especially for testers using IBM's APIs for testing, we presented a new definition of test cases and a test oracle in \method. The \method's generated test cases are directly compatible with IBM APIs, such as the Estimator API, which is heavily used for optimization and search circuits. We utilize the commuting property of Pauli strings to reduce the need for large number of predefined expected outputs and provide a new test oracle compatible with error mitigation methods, particularly those implemented in IBM APIs. Our empirical evaluation shows that \method can be a potential alternative for existing quantum circuit testing methods, improving the reliability and scalability of quantum computing methods developed using IBM's quantum computers.

\subsection{Integration with Other Quantum Software Frameworks}
\begin{figure}[!tb]


\includegraphics[width=\columnwidth]{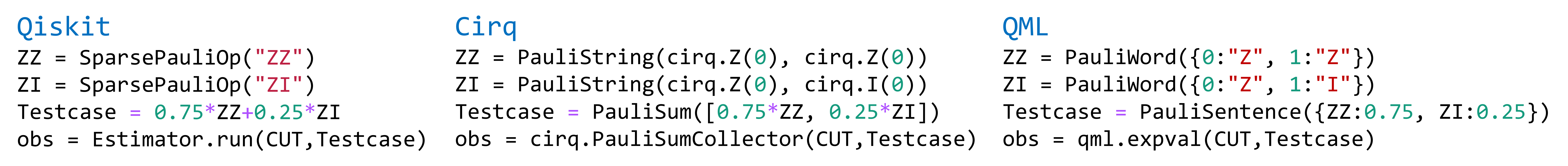}
\caption{An example of a test case in three widely used quantum software stacks: IBM's Qiskit, PennyLane's QML, and Google's Cirq.}
\label{fig:stack}
\end{figure}

Our approach demonstrates that the test case definition and oracle of \method are fully compatible with IBM's Qiskit framework. While our current implementation primarily supports IBM's API for automated test assessment, our test case definition and oracle are also compatible with other widely used gate-based quantum computers and software frameworks. These include IBM's Qiskit, PennyLane's QML, and Google's Cirq~\cite{qc_libraries}. Although the execution of quantum circuits varies among these frameworks, the fundamental definition of the test case and oracle remains consistent. For instance, consider a test case $\testCase = 0.75*ZZ + 0.25*ZI$, comprising two Pauli strings: \textit{ZZ} and \textit{ZI}. Figure~\ref{fig:stack} illustrates how this test case can be represented in the three most popular frameworks: Qiskit, QML, and Cirq.
All three frameworks support the concept of Pauli observables, enabling straightforward translation of the mathematical representation of test case \testCase into an executable form. In Qiskit, the test case can be instantiated using the SparsePauliOp object; in QML, using the PauliWord and PauliSentence objects; and in Cirq, using the PauliString and PauliSum objects. Each framework provides an API that accepts the test case and the circuit under evaluation, returning expectation values accordingly. Regarding \method's test oracle, no transformations are necessary for different frameworks, because the oracle is computed directly from the mathematical representation of the test case. Therefore, quantum software developers using various quantum software stacks can seamlessly integrate \method into their development environments.

\subsection{Scalability}
In terms of test assessment for quantum circuits, \method is a highly scalable method compared to existing testing methods. \method is capable of generating and evaluating $K \times N$ test cases from just $K$ Pauli strings specified in \ourPS, where $K$ represents the number of commuting Pauli families for a quantum circuit, and $N$ denotes the number of random test cases. Existing testing methods, which require a full program specification to determine the passing and failing of test cases, are limited to evaluating only the number of test cases explicitly specified in the program specification. In contrast, \method leverages the specified Pauli strings in \ourPS to generate as many test cases as the test budget allows, making it a practical approach for scenarios where the program specification is limited.

\subsection{Way Forward}
In this study, we introduced an alternative approach for quantum circuit testing by developing a new definition for test cases and a new test oracle. We demonstrated that quantum circuit faults can be effectively identified through randomly generated test cases. However, generating random test cases is time-consuming and resource-intensive, making it a highly inefficient method for testing. 

To enhance the efficiency and effectiveness of our method, we plan to improve the test case generation aspect of \method in the future. One promising solution is to employ search-based approaches such as genetic algorithms, simulated annealing, or particle swarm optimization~\cite{HarmanACMsurvery2012}. These algorithms can systematically explore the space of potential test cases, prioritizing those that are more likely to uncover faults efficiently. By using these search methods, we hope to significantly reduce the time and resources required for test case generation.

Another potential research direction is to discover relationships between test cases and various types of faults. Specifically, focusing on the relation between the passing or failing of tests within the same Pauli family. For instance, if one test in a family fails, is it more likely that other tests in the same family will also fail? Moreover, even if a test does not fail, could its execution provide a hint that another test in the same family might fail? Understanding the relation between specific test cases and the faults they reveal could lead to new methods for debugging and repairing quantum circuits. This knowledge would enhance the accuracy of fault detection and provide valuable insights for developing more robust quantum circuits.

\section{Related Work}\label{sec:qse}
Several studies have explored quantum software testing, introducing various methods for testing quantum circuits, evaluating test suite quality, and providing test oracles. Various tools and approaches have been proposed in the literature for testing quantum circuits. Quito~\cite{quitoASE21tool} is a tool proposed to test quantum circuits by fulfilling input-output coverage criteria. Tools like Muskit~\cite{mutation} and QMutpy~\cite{Rui_mutation} assist in mutation testing. QuSBT~\cite{search} uses a search algorithm to generate test cases and MuTG~\cite{mutation} us a multiobjective search algorithm to find a minimum number of test cases that can kill a maximum number of mutants. QuCAT~\cite{Combinatorial} employs combinatorial testing methods, while QuanFuzz~\cite{fuzz} utilizes fuzzing method for quantum circuit testing. Several methods have been developed to gauge the effectiveness of test suites. Coverage criteria such as input-output coverage~\cite{coverage} and equivalence class partitioning~\cite{LongTOSEM2024} are employed to evaluate test suite effectiveness. Currently, two common types of oracles are used for quantum circuit testing. The first type employs statistical tests to compare quantum circuit outputs~\cite{coverage, Huang2019}. The second type, property-based testing~\cite{property,property2}, derives several properties for the circuit under test, where the violation of these properties indicates the presence of faults. Several bug identification frameworks~\cite{bug1,bug2,MiranskyyICSE2020} have also been proposed, including Bugs4Q~\cite{bug3}, which is a benchmark for real-world Qiskit circuit bugs.

Current real-world quantum computers are noisy, which greatly reduces the reliability of computations and introduces errors in circuit outputs~\cite{noise}. The causes of noise in quantum computers are environmental factors such as magnetic fields and radiation, qubit cross interactions, and calibration error in quantum gates~\cite{noisesource}. Due to noise, testing quantum circuits on actual quantum computers is a challenging task because it becomes difficult to differentiate between faults and errors from noise. Several methods have been proposed for mitigating quantum noise~\cite{qem}, focusing on reducing noise effects on the expectation values of quantum circuits. To the best of our knowledge, only three testing methods enable quantum circuit testing on real computers. The first method is QOIN~\cite{qoinArxiv} which trains a machine-learning~(ML) model utilizing three statistical features to reduce the effect of noise. The second is Qraft~\cite{qraft}, which introduced a new set of features for training ML models that can be applied to multiple circuits under test. The last one is Q-LEAR~\cite{qlearFSE2024}, which introduced a new feature called depth-cut-error to address Qraft's limitation of overestimating noise in the circuit output.

The main drawback of current testing methods is their incompatibility with search and optimization circuits, as well as the necessity for full program specification. Apart from three methods~\cite{qoinArxiv,qlearFSE2024,qraft}, existing testing approaches are not suitable for use with real noisy quantum computers and depend on statistical test-based oracles that are incompatible with existing error mitigation methods. Even the three methods that do support testing on real quantum computers still face the issues of incompatibility with search and optimization circuits and the requirement for full program specification.

\section{Conclusion}
To address the practical challenges of incompatible test case definition, costly requirement of full program specification, and incompatibility with error mitigation methods, we developed \method, which is a novel quantum circuit testing approach. \method utilizes Pauli strings as a new definition of test cases, eliminating the need for having explicit inputs in quantum circuits and ensuring compatibility with industrially relevant optimization and search circuits. We have also introduced a new test oracle that can be easily integrated with commonly adopted error mitigation methods in the industry. Furthermore, by utilizing the commuting property of Pauli strings, we also reduce the cost of a full program specification making quantum software testing more practical in industrial settings. We evaluated \method on 194,982 quantum circuits, benchmarking it against the current state-of-the-art test case definition and test oracle. The results demonstrate the effectiveness of \method in terms of test assessment, achieving perfect F1-score, precision, and recall. Additionally, we validated the industrial applicability of \method by conducting test assessments on IBM's three real quantum computers, leveraging both industrial-grade and open-source error mitigation methods. In the future, we aim to enhance the applicability of \method in industrial settings by integrating \method with search-based approaches to further reduce the time and resources required for test assessment.

\subsection*{Acknowledgments}
The work is supported by the Qu-Test project (Project \#299827) funded by the Research Council of Norway. S. Ali is also supported by Oslo Metropolitan University's Quantum Hub. Paolo Arcaini is supported by Engineerable AI Techniques for Practical Applications of High-Quality Machine Learning-based Systems Project (Grant Number JPMJMI20B8), JST-Mirai.

\bibliographystyle{ACM-Reference-Format}

\end{document}